\documentclass{PoS}
\usepackage{subfigure}
\title{A Stochastic Method for Semileptonic Form Factor Calculations on the Lattice}

\ShortTitle{A Stochastic Method for Semileptonic Calculations}

\author{\speaker{Richard Evans}, Gunnar Bali, Sara Collins\\
        Institute for Theoretical Physics ,University of Regensburg\\
        93053 Regensburg, Germany \\
        E-mail: \email{richard.evans@physik.uni-r.de}}

\abstract{We investigate an alternative to the Sequential
Propagator Method used in Lattice QCD calculations of
semileptonic form factors. We replace the sequential propagator with a
stochastic propagator so that, in principle, all momentum and sink smearing
combinations are available with only a single spin-color inversion.
Practically, the stochastic noise is significant and must be reduced
at the cost of more inversions.
We study the behavior of the stochastic noise and compare the computational costs of this stochastic technique and the Sequential Propagator Method.  We also present preliminary
semileptonic form factor results using the stochastic technique on $N_f=2$
configurations with a non-perturbatively improved Sheikoleslami-Wohlert action generated by the QCDSF collaboration.  At a fixed cost, measured in terms of the number of heavy-quark inversions, the method provides more correlators for the extraction of the form factors at various $q^2$'s than the Sequential Propagator Method.  These additional correlators reduce the total statistical errors of certain kinematic points, although the stochastic error is still comparable to the gauge error at other points.}

\FullConference{The XXVII International Symposium on Lattice Field Theory - LAT2009\\
		 July 26-31 2009\\
		 Peking University, Beijing, China}

\begin{document}
\graphicspath{{../}}
\section{Introduction}
Experimental measurements of heavy-light semileptonic decays, combined with theoretical input,  can be used to extract the Cabbibo-Kobayashi-Maskawa (CKM) matrix elements $|V_{ub}|$, $|V_{cb}|$, $|V_{cd}|$, and $|V_{cs}|$.  The determination of these matrix elements provides constraints on the CKM Unitarity Triangle and thus test the Standard Model.  Conversely, $|V_{cd}|$ and $|V_{cs}|$ are known with high precision, and can be used to test the Lattice techniques used to calculate the decay rates of $D$ mesons. 
The approach to current Lattice calculations of the semileptonic decay rates of $D$ mesons involves constructing the appropriate three-point function using the so-called sequential (or extended) propagator.  We investigate an alternative method using stochastic techniques, which we refer to as the Stochastic Sink Method (SSM), in the hope of achieving an overall savings in computational effort.  With this method all momentum and sink smearings are in principle available using only a single spin-color inversion.  In this report we present a basic comparison at fixed cost of the two methods, and provide preliminary results of the form factors using the SSM.   

In the following we focus on the decays of a heavy-light pseudoscalar charm-like meson ($H=D$) to a light-light pseudoscalar meson ($P=\pi/K$) and leptons ($l,\nu_l$).  For these processes the differential decay rate can be parametrized as
\begin{equation}
\frac {d \Gamma}{dq^2}(H\to P l\nu_l)=\underbrace{|V_{cd/s}|^2}_{CKM}\underbrace{\frac{G_F^2}{192\pi^2m^3_{H}}\lambda^{3/2}(q^2)}_{Perturbatively \, known}\underbrace{|F_{+}(q^2)|^2}_{form \, factor},
\end{equation}
where $q^2=(p_H-p_P)^2$ is the squared difference between the initial and final state four-momentum.  The greatest source of uncertainty in the theoretical calculation is due to the non-perturbative interactions parametrized by the form factor $F_+(q^2)$.

  These interactions appear in the hadronic matrix element $\langle P(p_P)|V_{\mu}(q^2)| H(p_H)\rangle$. $V_{\mu}=\bar{\psi_c}\gamma_{\mu}\psi_l$ is a weak flavour-changing vector current, where $\psi_c$ is the charm quark and $\psi_l$ is the $d$ or $s$ quark.
 The matrix element can be parametrized as a linear combination of the form factors $F_+$ and $F_0$,
\begin{equation}
\langle P(p_P) |V_{\mu}(q^2)| H(p_H)\rangle=\left\{ p_H+p_P-q(m_H^2-m_P^2)/q^2\right\}_{\mu}F_+(q^2)+\left\{ q(m_H^2-m_P^2)/q^2\right\}_{\mu} F_0(q^2).
\end{equation}  
On the lattice the matrix elements are extracted from three-point correlators with the following form,
\begin{eqnarray}\label{eq:c3}
C_3(T,t;\vec{p}_H,q)=\sum_{\vec{x},\vec{y}}e^{-i\vec{p}_H\cdot \vec{x}}e^{\vec{q}\cdot\vec{y}}\langle 0|\bar{\psi}_u\gamma_5\psi_c(\vec{x},T)\cdot \bar{\psi}_c\gamma_{\mu} \psi_l(\vec{y},t)\cdot \bar{\psi}_l \gamma_5 \psi_u(\vec{0},0)|0\rangle= \nonumber \\
-\sum_{\vec{x},\vec{y}}e^{-i\vec{p}_H\cdot \vec{x}}e^{\vec{q}\cdot\vec{y}}\mathrm{Tr} \left< M^{-1}_u(\vec{0},0;\vec{x},T))\gamma_5 M^{-1}_c(\vec{x},T;\vec{y},t)\gamma_{\mu}M_l^{-1}(\vec{y},t;\vec{0},0)\gamma_5\right>
\end{eqnarray}
where $\psi_u$ is the spectator light-quark and $M^{-1}_x$ is the propagator for quark $x$.  In the limit of large time separation Eq.~(\ref{eq:c3}) has the form
\begin{equation}
\mathop {\lim} \limits_{T\gg t \gg0} C_3(T,t;\vec{p}_H,q)\to \frac {Z_P}{2E_P} \frac{ Z_H}{2E_H}\times \langle P(p_P)|V_{\mu}|H(p_H)\rangle  \times e^{-E_P t} e^{-E_H(T-t)},
\end{equation}
so that a determination of the amplitudes and energies from ratios or simultaneous fits with meson propagators can isolate the matrix element.
\section{ Stochastic Sink Method (SSM)}
The standard method for calculationing $D$ meson semileptonic three-point functions uses sequential propagators.  The sequential propagator provides a way to calculate a heavy-quark propagator that connects all spatial sites $\vec{x}$ at the sink time-slice $T$, to all sites $\vec{y}$ and $t$ at the vector current.  It starts by taking a single time-slice of the spectator quark propagator, $M^{-1}_u(\vec{x},T;\vec{0},0)$.  The  desired sink momentum $p_H$, sink smearing $W_s$, and sink gamma $\Gamma_f$ are then inserted to get the ``sequential source". The heavy-quark action is then inverted on this ``sequential source" to get the sequential propagator,
 \begin{equation}   
\sum_{\vec{x}}M^{-1}_c(\vec{y},t;\vec{x},T) e^{i\vec{p}_H\cdot \vec{x}} \Gamma_f W_s M^{-1}_u(\vec{x},T;\vec{0},0). 
\end{equation}
The sequential propagator can then be combined with the daughter light-quark propagator, $M_l(\vec{y},t;\vec{0},0)$, and appropriate gamma matrices to get Eq.~(\ref{eq:c3}).
This method requires a heavy-quark inversion for each distinct sink momentum $p_H$, sink smearing $W_S$, and sink gamma $\Gamma_{f}$.  The computational effort required for this procedure can become prohibitive if many sink momenta and/or smearings are needed, as would be required in the use of the Variational Method \cite{Michael:1985ne,Luscher:1990ck} for studying excited state decays.    

An alternative method which may be more efficient is to replace the sequential propagator with an all-to-all propagator \cite{Bernardson:1993yg}.   We construct all-to-all propagators by generating random vectors $\eta^{[r]}_j(\vec{x},T)$, $r=1,\ldots,N$, at a particular timeslice $T$ using complex $Z_2$ noise, 
 with the property 
\begin{equation}
 \frac 1 N \sum_r \eta^{[r]}_i(x) \eta^{\dagger [r]}_j(z)=\delta_{xz}\delta_{ij}+{\cal O}(1/\sqrt{N})
\end{equation}
 where $i,j$ label spin and color.  We then invert the charm-like quark's Dirac operator, $M_{ij}(x,y)$, on each source $\eta_j^{[r]}(\vec{x},T)$ to obtain the solutions $\psi_j^{[r]}(\vec{y},t)$:
\begin{equation}
M_{kj}(\vec{z},T;\vec{y},t)\psi^{[r]}_j(\vec{y},t)=\eta^{[r]}_k(\vec{z},T)\to \psi^{[r]}_j(\vec{y},t)=\sum_{\vec{z},k}M^{-1}_{jk}(\vec{y},t;\vec{z},T)\eta^{[r]}_k(\vec{z},T).
\end{equation}
The average over the product of the sources and solutions provides an estimate for the all-to-all heavy-quark propagator,
\begin{equation}
\frac 1 N \sum_r \psi^{[r]}_j(\vec{y},t) \eta^{\dagger [r]}_i(\vec{x},T) =\underbrace{ M^{-1}_{ji}(\vec{y},t;\vec{x},T) }_{ \rm{all-to-all } }+\sum_{\vec{z},k}M^{-1}_{jk}(\vec{y},t;\vec{z},T)\underbrace{(\frac 1 N \sum_r \eta^{[r]}_k(\vec{z},T)\eta^{\dagger [r]}_i(\vec{x},T)-\delta_{\vec{z}\vec{x}}\delta_{ki})}_{\rm{ error}\propto{\cal O}(1/\sqrt{N})},
\end{equation}
where the stochastic error decreases with $N$, the number of source/solution pairs used.  

A stochastic estimate of Eq.~(\ref{eq:c3}) can be constructed by combining the spectator and daughter point-to-all light quark propagators and the all-to-all heavy-quark sources and solutions in the following manner,
\begin{eqnarray}
& -\frac 1 N \sum_r \mathrm{Tr} \left<\sum_{\vec{x}} e^{-i\vec{p}_H\cdot \vec{x}}  \Gamma_i M_u^{-1}(\vec{0},0;\vec{x},T)\Gamma_f \eta^{[r]}(\vec{x},T)\cdot  \sum_{\vec{y}}e^{\vec{q}\cdot\vec{y}} \psi^{\dagger}(\vec{y},t)\Gamma M_l(\vec{y},t;\vec{0},0) \right> =   \nonumber \\ & C_3(T,t;\vec{p}_H,q) + {\cal O}(1/\sqrt{N}), & 
\end{eqnarray}
where the appropriate propagator smearings must be applied and $\Gamma_f=\Gamma_i=\gamma_5$ and $\Gamma=\gamma_{\mu}$. 

If the stochastic error term were negligible compared to the gauge noise we would have all sink smearings and momentum available with a single heavy-quark inversion.  For the parameters in our calculation the error term is not neglible and must be reduced by additional noise vectors and/or noise reduction methods.  Both improvements involve a computational overhead which must be accounted for in any comparison between the SSM and Sequential Propagator Method.  We tried all combinations of spatial even/odd, color, and spin partitioning \cite{Bernardson:1993yg}, and found spin partitioning on its own to be the most computationally efficient method of noise reduction.  The number of vectors used will be discussed in the next section. 
\section{Sequential Propagator Method versus SSM and Preliminary Results}
We perform a simple comparison of the two methods making basic assumptions about the data set desired for our form factor calculation.  We compare the total statistical errors of correlators constructed from the two methods at fixed cost, where the cost is measured in the number of heavy-quark action inversions.  All available rotationally equivalent correlators corresponding to each individual $q^2$ point  are averaged over to improve the statistics.  The sink is placed at the midpoint of the lattice, allowing us to fold the data along the time axis.

\begin{table}
\begin{center}
 \begin{tabular}{|ccc|}
\hline
Param & $16^3\times 32$ & $24^3\times 48$ \\
\hline 
$n_{cfgs}$ & 100 & 221 \\
$a$ & 0.089 fm & 0.076 fm\\
$m_{\pi,sea}$ & $929$ MeV & $272$ MeV\\
$m_{\pi,valence}$  & $929$ MeV & $438$ MeV\\
$m_{D}$ & $2.44$ GeV& $1.93$ GeV\\ 
\hline
\end{tabular}
\caption{Parameter details for the 2 ensembles used in the cost comparison.}\label{tab:params}
\end{center}
\vspace{-0.75cm}
\end{table}
Two QCDSF ensembles with 2 dynamical sea-quarks  \cite{Ali Khan:2003cu} were used in the comparison with the parameters shown in Tab.~\ref{tab:params}.  The ensembles were generated using a Wilson Plaquette action for the gluons and a non-perturbatively improved  Sheikoleslami-Wohlert action for the fermions.  Both actions have errors starting at ${\cal O}(a^2)$. The interpolating fields in the correlators were Wuppertal smeared \cite{Gusken:1989ad}, using the smearing factor and number of iterations that optimize overlap with the light-light meson ground state.  It should be noted that the number of smearing iterations increases the stochastic noise significantly faster than it increases the gauge noise, and thus using a more customized smearing (with less iterations) for the heavy-light state $H$ would reduce the stochastic noise presented in this report.

We assume four sink momenta are desired: $p_H=(0,0,0),(1,0,0),(1,1,0),(1,1,1)$.  These sink momenta create data in the physical region, $q^2>0$, for the meson masses used in our calculation.  We also assume a minimal smearing basis of two different sink smearings, $W_s$.  The cost required to build this data set using the Sequential Method is thus $12(spin/color) \times 4 (p_H) \times 2(W_s)=96$. The cost for the SSM with spin partitioning is $4(spin)\times N$, so using $N=24$ is of comparable cost to the Sequential Method.  Note that all $p_H$ are generated at negligible cost with the SSM. 

\begin{figure}
\subfigure[] {\label{fig:a}  \includegraphics[trim = 2cm 0 0cm 0,clip=true width=7cm,height=6.5cm]{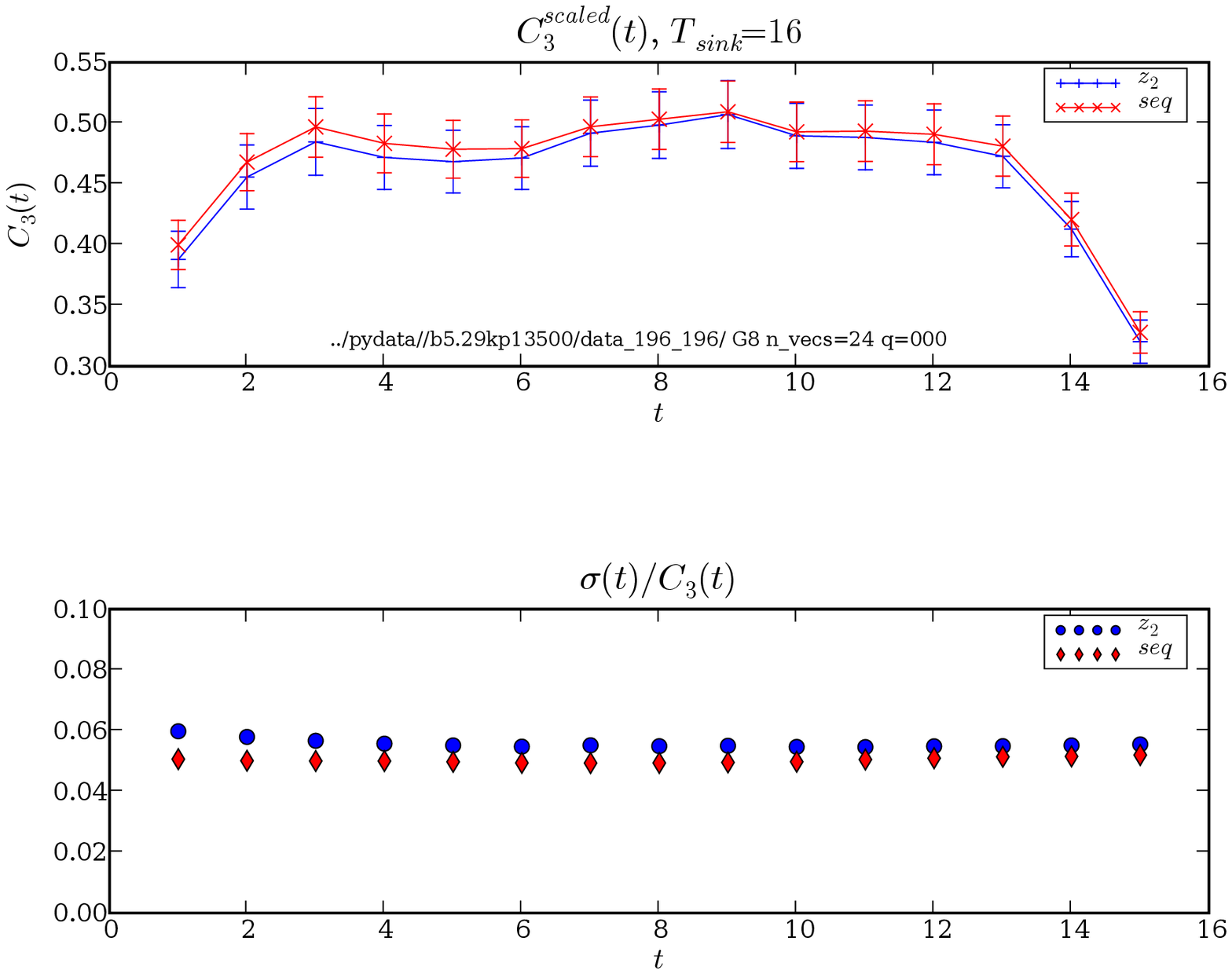}}
\subfigure[] {\label{fig:b} \includegraphics[trim =30 0 0 0,clip=truewidth=7cm,height=6.5cm]{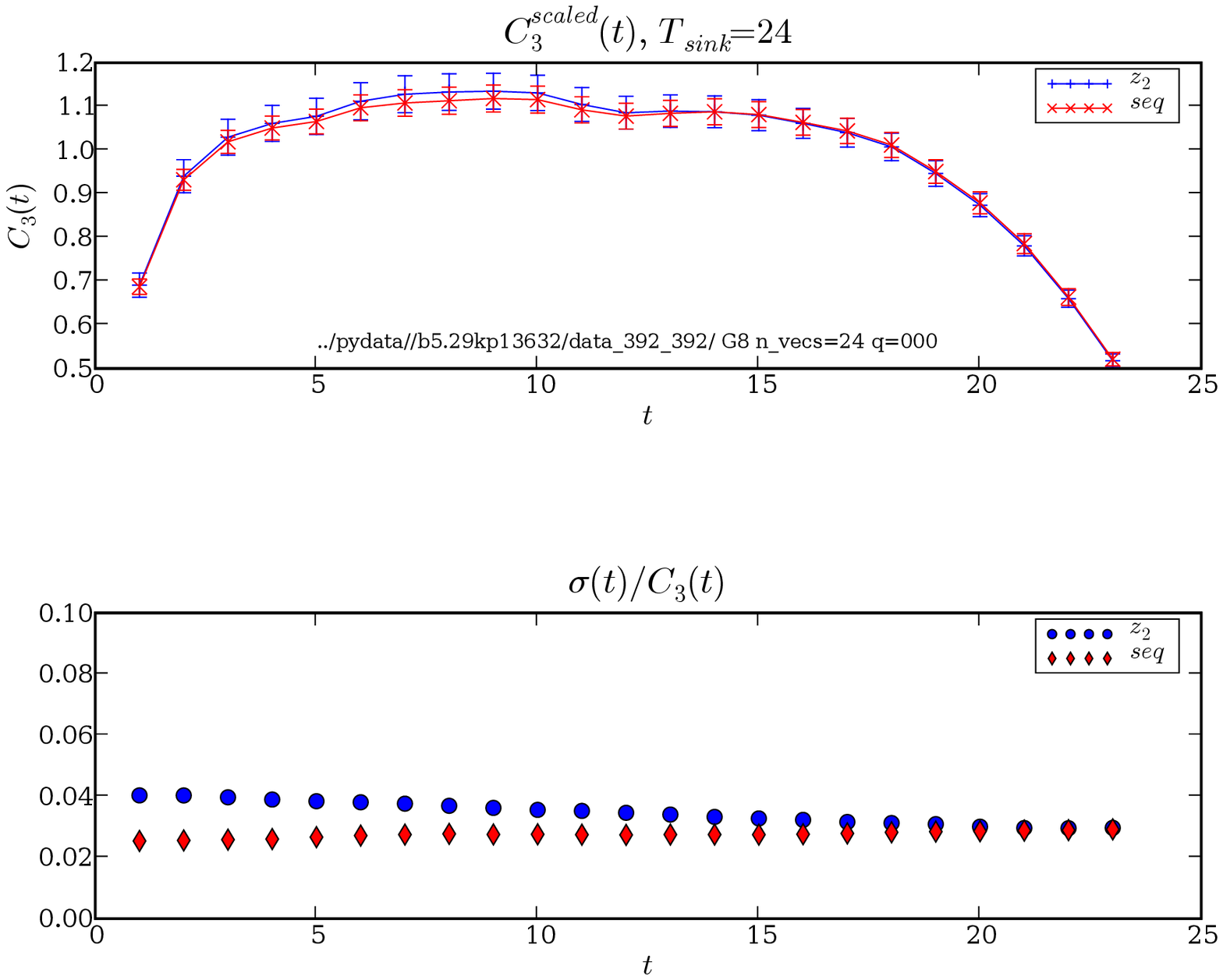} }\\
\subfigure[] {\label{fig:c} \includegraphics[trim = 2cm 0cm 0cm 0cm,clip=true width=16cm,height=6.5cm]{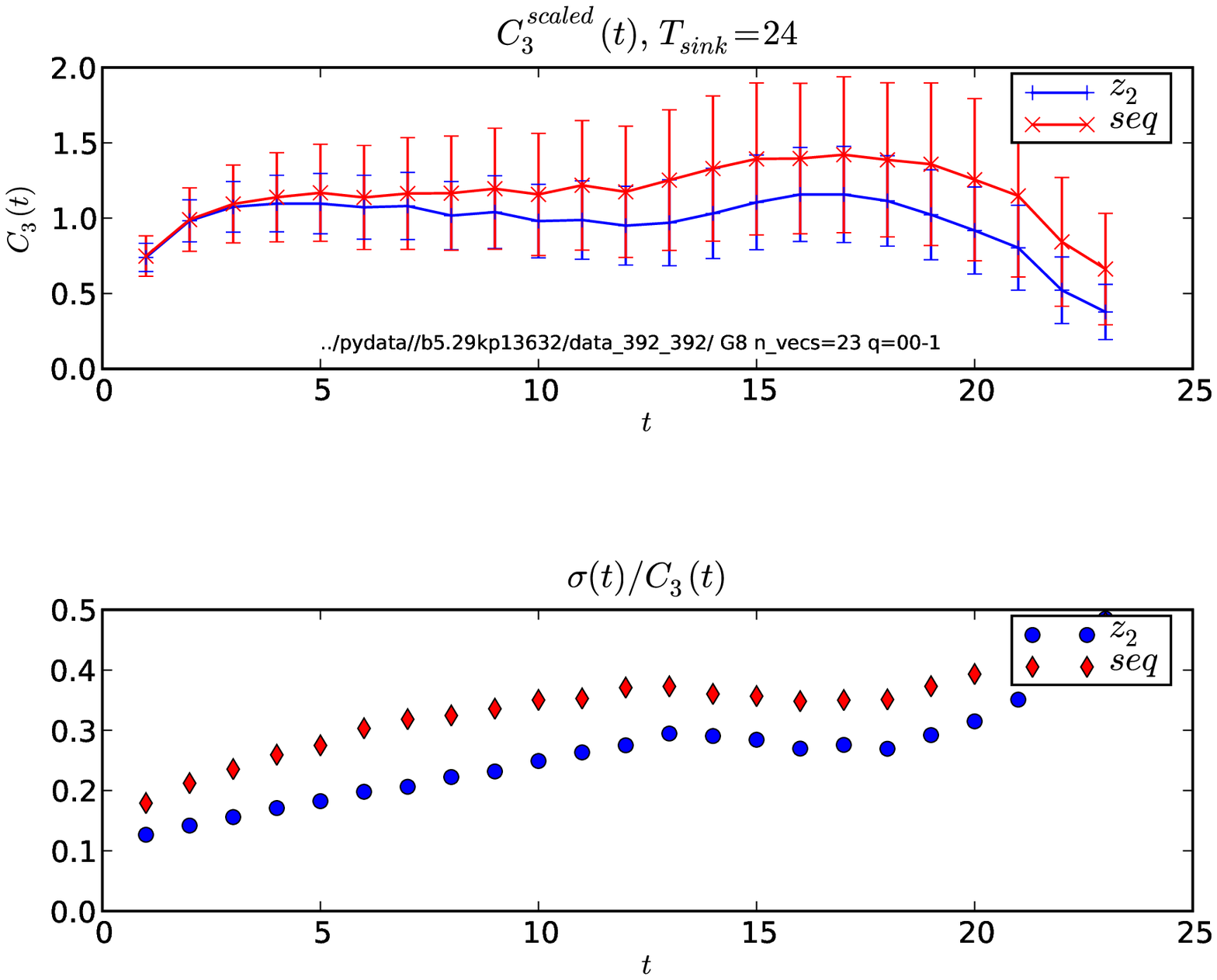}}
\subfigure[] {\label{fig:d} \includegraphics[trim = 30 0cm 0cm 0cm,clip=true width=16cm,height=6.5cm]{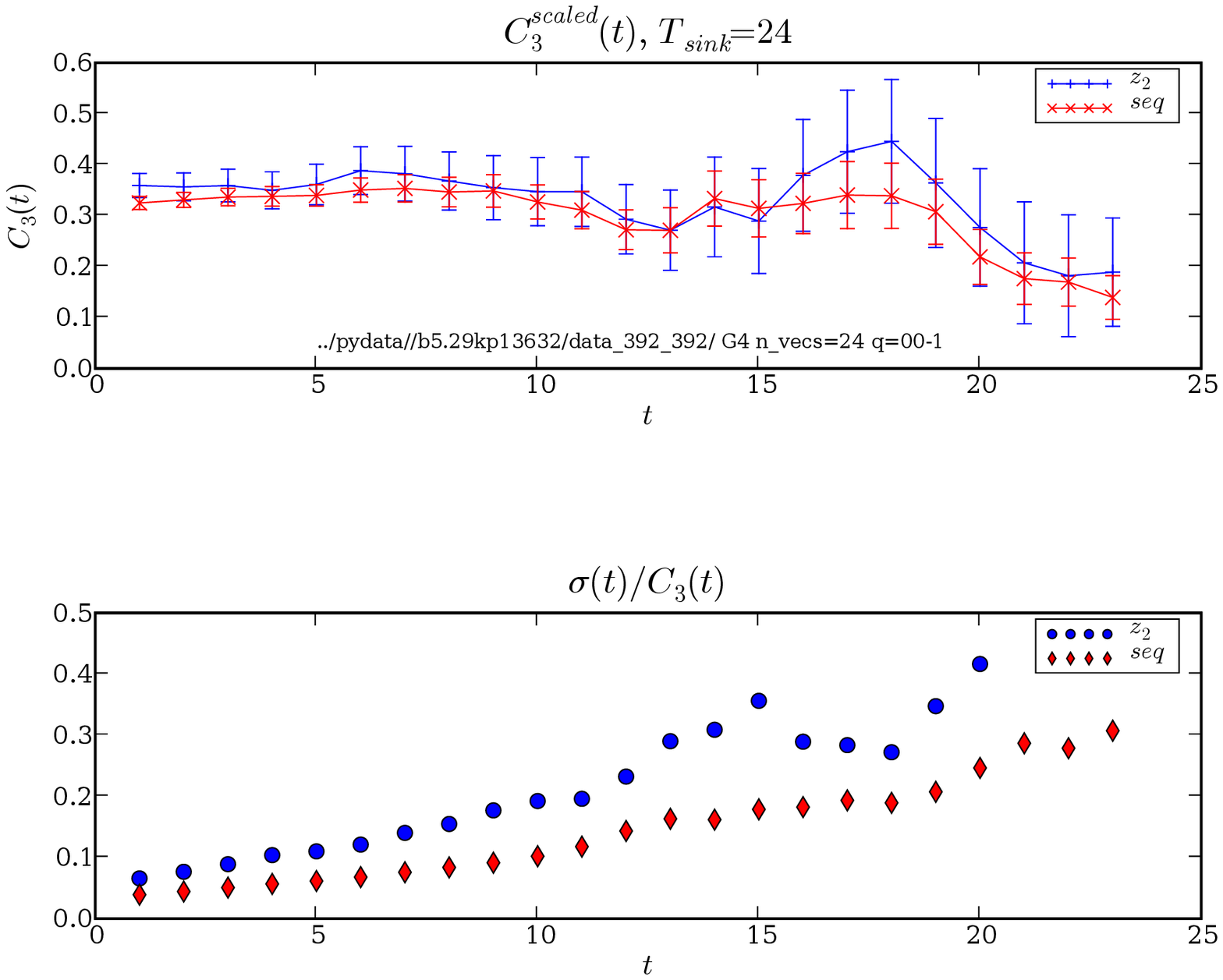}}
\caption[]{Four correlators corresponding to different $q^2$ kinematic points are shown.  The upper plot of each figure shows the sequential (red) and stochastic (blue) correlators scaled by the appropriate two-point functions, such that the vector matrix element is proportional to the resulting plateau.  The lower plot of each figure shows the percentage statistical errors of the two methods.  The correlators shown are chosen because they are representative of the behaviour of the statistical errors in our data set.
In Fig.~\ref{fig:a} and Fig.~\ref{fig:b} the temporal component of the vector current at $q^2=q^2_{max}$ for the $16^3\times 32$ and $24^3\times 48$ ensemble are shown.  In Fig.~\ref{fig:c} the  temporal component with $|p_{P}|=1$ and $|p_{H}|=1$ is shown, where the stochastic correlator has been constructed from the average of the six rotationally equivalent correlators.  The Sequential Method has only one correlator available in this data set.  Fig.~\ref{fig:d} presents the spatial component of the matrix element, where $|p_{P}|=1$ and $|p_{H}|=0$.  Both the sequential and stochastic correlators have been averaged over the six available rotationally equivalent correlators. } 
\end{figure}

Figs.~\ref{fig:a}-\ref{fig:d} are representative of the range of behaviour of the percentage statistical errors in our data set using $N=24$. Figs.~\ref{fig:a} and \ref{fig:b} show the $q^2=q^2_{max}$ kinematic point for the $16^3\times 32$ and $24^3\times 48$ lattices respectively.  At this kinematic point the initial and final state mesons have zero spatial momentum.  The total noise for the correlator on the $16^3\times 32$ lattice shown in Fig.~\ref{fig:a} is dominated by the gauge noise-in fact this is true for all other kinematic points on this lattice.  The stochastic noise for the $24^3\times 48$ lattice shown in Fig.~\ref{fig:b} is dominated by the gauge noise near the heavy-light meson ($t=24$), but becomes comparable to the gauge noise at the light-light meson ($t=0$). 

The SSM correlator shown in Fig.~\ref{fig:c} is constructed from an average of all available rotationally equivalent correlators, 6 in total, whereas the Sequential Method generated correlator has only a single correlator available.   The averaging causes the SSM correlator to actually have smaller errors than the Sequential Method generated correlator.  This suggests that there is statistical gain to be had in so many additional sink momenta.  Fig.~\ref{fig:d} shows the noisiest SSM generated correlator, relative to the Sequential Method correlator, that we found in our data set.  In this case both the SSM and Sequential Method correlators can be constructed from 6 rotationally equivalent correlators.  

We have examined all correlators with an appreciable signal and draw the conclusion that with the modest number of 24 stochastic estimates the stochastic error is at worst comparable in magnitude to the gauge noise, while at best the additional data available results in smaller total errors for the SSM.  We are expanding this analysis and expect to have more quantitative results in the near future.  

\begin{figure}
\begin{center}
\includegraphics[trim = 1cm 1cm 1cm 1cm,clip=true width=16cm,height=5cm]{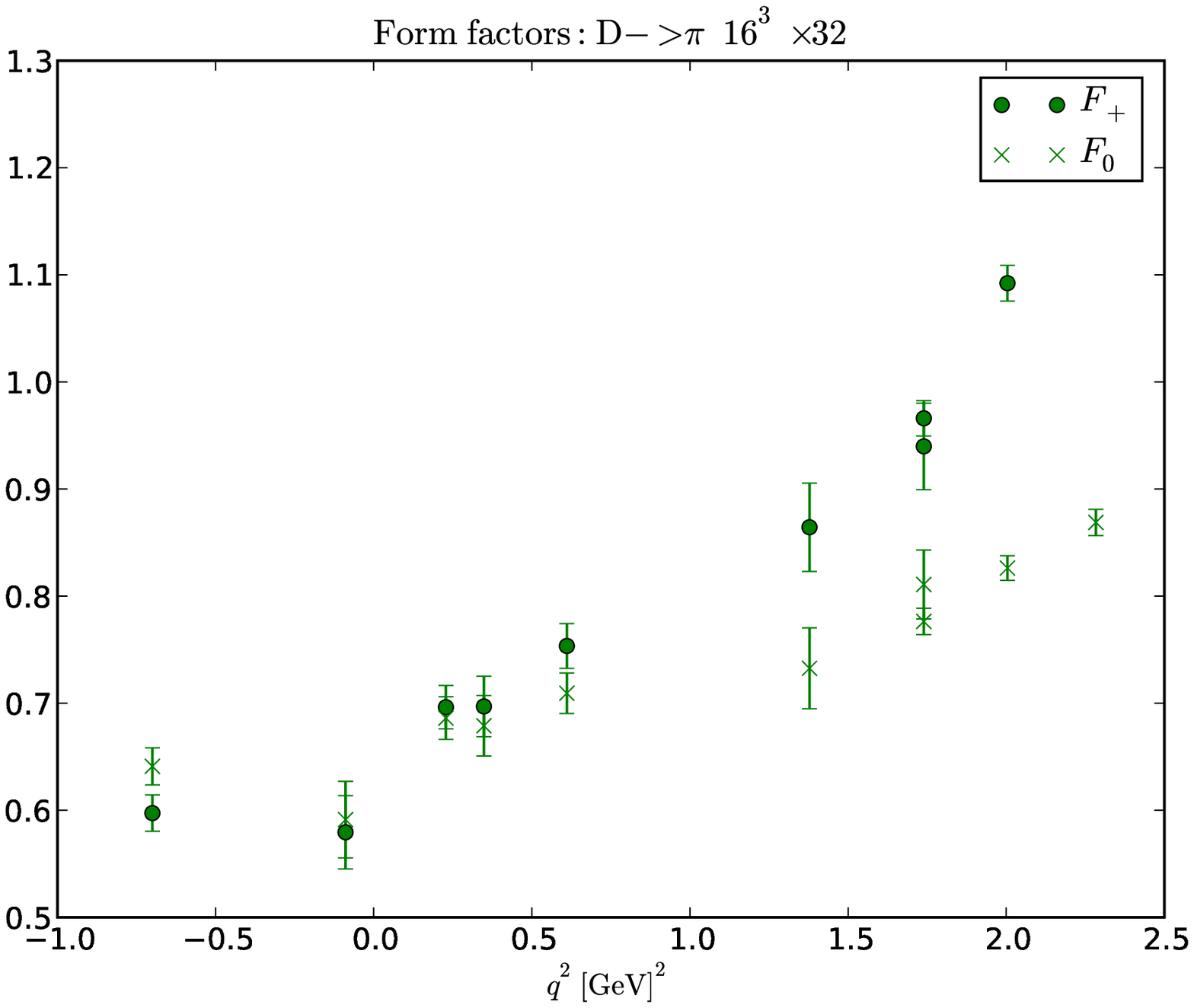}
\caption{Preliminary results using the SSM on the $16^3\times 32$ ensemble with $N=24$ and 600 configurations.} \label{ff24}
\end{center}
\vspace{-0.75cm}
\end{figure}
Bare lattice results using the SSM for $F_+(q^2)$ and $F_0(q^2)$ calculated from the $16^3\times 32 $ ensemble are presented in Fig.~\ref{ff24}.  600 configurations with 24 stochastic vectors were used with all rotationally equivalent correlators averaged to improve statistics.  
\section{Additional Considerations and Outlook}

The effects of ${\cal O}(a)$ improvement have also been examined and do not change our conclusions.  The matching coefficient $Z_V$ is known non-perturbatively  for these lattices \cite{Bakeyev:2003ff} and the coefficient of the improvement term $c_V$ is known to one-loop perturbatively \cite{Sint:1997jx}.  

We've also investigated using the ``one-end trick''\cite{Foster:1998vw,McNeile:2006bz} with one stochastic vector, with and without spin partitioning.  In our examination, comparing the three methods with no momentum averaging, we saw a noise reduction for certain $q^2$'s in the one-end generated correlators.  
For all $q^2$'s however, after averaging over the available correlators for the SSM and Sequential Method (and no averaging for the one-end correlators), the one-end correlators' errors were larger. Because generating the additional rotationally equivalent correlators or stochastic vectors for the one-end method would make it clearly more expensive than the other methods, we've concluded the one-end method is less efficient for the parameters we are working with.  

The SSM is potentially computationally more efficient than using sequential propagators.  With the SSM all sink momentum and smearings can be generated with a fixed number of heavy-quark inversions.  Whether this method saves computational effort over the Sequential Method depends on the statistical improvement that additional sink momenta and smearings provide for the extraction of the form factors, and a more comprehensive investigation is underway.   At this stage it can be stated that at a fixed cost the errors at particular $q^2$'s can be reduced by using the SSM.  This method should also significantly reduce the cost required to use the Variational Method in three-point calculations, where multiple sink smearings are required for each sink momentum.  

 \acknowledgments
 The Chroma software suite \cite{Edwards:2004sx} was used extensively in this work.  The gauge configurations are provided by the QCDSF collaboration via the ILDG.  The simulations were run on the Athene Cluster of the University of Regensburg.  Our work is supported by the DFG Sonderforschungsbereich/Transregio 55.  Sara Collins acknowledges support from the Claussen-Simon-Foundation (Stifterband f\"ur die Deutsche Wissenchaft).

\end{document}